\documentstyle[11pt,graphicx]{article} 
\begin{document} 

\title{Sensitivity optimization of multichannel searches for new signals} 
\author{Giovanni Punzi\\ 
\it Universita' di Pisa and INFN-Pisa
} 
\maketitle

\begin{abstract}

The frequentist definition of sensitivity of a search for new phenomena proposed in \cite{Punzi:2003bu} has been utilized in a number of published experimental searches. In most cases, the simple approximate formula for the common problem of Poisson counts with background has been deemed adequate for the purpose. There are however many problems nowadays in which more complex analysis is required, involving multiple channels. In this article, the same approach of ref. \cite{Punzi:2003bu} is applied to a multichannel Poisson problem, and a convenient formula is derived in closed form, generalizing the known result for the simple counting experiment. An explicit solution is also derived for the common case of a search for a Gaussian signal superimposed over a flat background.

\end{abstract}

\section{Introduction} \label{sec:Intro}

The approach described in ref. \cite{Punzi:2003bu} can the briefly summarized in this way: there is an experiment in which we want to test a certain $H_0$ taking the form of $\mu=0$ against an alternative $\mu \ne 0$, at a predetermined value of significance $\alpha$; if the test fails to reject $H_0$, we additionally want to set Confidence Limits on $\mu$ with some chosen confidence level CL. For reasons explained at length in that paper, it is proposed that optimization of the experimental procedure should be based on maximizing the set of values of $\mu$ such that:
\begin{equation}\label{eq:sens}
1-\beta_\alpha(\mu)>CL
\end{equation}

This region of $\mu$ can be thought of as a region of parameters to which the experiment is ``sufficiently sensitive". 

\section{Case of the simple counting experiment}\label{sec:examples}

In this case, we have the discrete observable $n$, the number of events observed, which is Poisson-distributed with a mean determined by $B(t)$, the expected number of background events (supposed known, given some parameters $t$ under the experimentalist's control), and the possible contribution of signal events $S_m(t) = \epsilon(t)\sigma(m)$, where $\sigma(m)$ is the physical cross section for production, given the parameter $m$:
\begin{eqnarray}
p(n|H_0) = &e^{-B(t)} B^n(t)/n!\\
p(n|H_m) = &e^{-B(t)-S_m(t)} \left[B(t)+S_m(t)\right]^n/n! 
\end{eqnarray}

For this problem, it is easy to see that the only sensible definition of the critical region for the test of $H_0$ takes the form 
$n>n_{min}$.
In that case one finds that, in a gaussian approximation, the maximum sensitivity is attained by maximizing the quantity (independent of $\sigma(m)$):

\begin{equation} \label{eq:completemax}
 \frac{ \epsilon(t)}{ b^2+ 2\,a\,{\sqrt{B(t)}} + 
  b\,{\sqrt{b^2 + 4\,a\,{\sqrt{B(t)}} + 4\,B(t)}}}
\end{equation}

 The expression simplifies considerably when $b\simeq a$, and becomes:

\begin{equation} \label{eq:simplemax}
 \frac{ \epsilon(t)}{ a/2 + \sqrt{B(t)}}
\end{equation}

\section{Generalization to multiple counting experiments}

Suppose now that the signal produces an increase in rate in $n$ independent bins. This may be the case of combining some individual measurements, or possibly of a histogram, where the signal would provide an additional contribution, of a {\em fixed shape} . The observation will be a set of $N$ integers $\{k_i\}$, and this is the observable data from which the experimental inferences will be made. Each bin is populated by a (known) background level $b_i(t)$, plus a possible signal  $s_i(m,t) = \epsilon_i(t) \mu_m$. 

Assuming independence of all bins, and a fixed total number of events N, we have the {\em pdf}:
\begin{equation}
p(\vec{k};\vec{\epsilon},\vec{s},\vec{b}) = \frac{N!}{\prod_i{k_i!}}\prod_i{(b_i+\epsilon_i \mu)^{k_i}}
\end{equation}

In the single-channel case there was an obvious test to use for verifying the null hypothesis; here things are a bit more complicated, and we need to first determine what an appropriate test could be.  Given that the work is aimed at a search for a signal that is expected to be, if any, just emerging from the background (this is the regime the whole approach is targeting), it makes sense pick the Fisher score function $S_F = \frac{\partial\log{\cal L}}{\partial \mu}$, evaluated in $\mu=0$, as a test statistic. This is well-known to yield a Locally Most Poweful one-sided test in the vicinity of $\mu=0$. In our case, the test statistics turns out to be:

$$ T(\vec{k}) = \frac{\partial}{\partial \mu} \left.\sum_i{ k_i\log{(b_i+\epsilon_i\mu)}}\right|_{\mu=0} = \sum_i{\frac{\epsilon_i}{b_i}k_i}$$

We obtained a quite simple test; what we need now is to find its critical region. The distribution of the test statistics $T$ is clearly that of a linear combination of Poisson; given that there are likely several terms in this sum, we will confidently take it to be Normal, with mean and variance:
\begin{eqnarray}
E[T] = \sum_i{\epsilon_i}\\
\textrm{Var}[T] = \sum_i{\frac{\epsilon_i^2}{b_i}}
\end{eqnarray}
The test for $H_0$ then takes the explicit form:
\begin{equation}
T- \sum_i{\epsilon_i} >a\sqrt{\sum_i{\frac{\epsilon_i^2}{b_i}}}
\end{equation}
where $a$ is, as usual, the number of gaussian sigmas corresponding to the chosen test size $\alpha$ (one-sided).

We now need to evaluate the power of this test. Still in the gaussian approximation, we obtain for the mean and variance of the test statistics under $H_m$:

\begin{eqnarray}
E[T] = \sum_i{\epsilon_i} + \left(\sum_i{\frac{\epsilon_i^2}{b_i}}\right)\mu\\
\textrm{Var}[T] = \left(\sum_i{\frac{\epsilon_i^2}{b_i}}\right) + \left(\sum_i{\frac{\epsilon_i^3}{b_i^2}}\right)\mu
\end{eqnarray}

The master equation (\ref{eq:sens}) then becomes:

\begin{equation}
A\mu \ge a\sqrt{A} + b\sqrt{A+ B\mu}
\end{equation}

where for compactness we defined the two quantities:
\begin{eqnarray}
A = \sum_i{\frac{\epsilon_i^2}{b_i}}\\
B = \sum_i{\frac{\epsilon_i^3}{b_i^2}}
\end{eqnarray}

and $b$ has its usual meaning of the number of gaussian sigma corresponding to CL.

Isolating the $b$ term, squaring, and rearranging we obtain:
$$ A^2\mu^2 - (2 a A^{3/2} + b^2B)\mu + (a^2-b^2) \ge 0$$

this quadratic can easily be solved for $\mu$ to yield an exact solution. However, it is simpler and more convenient to use here the same approximation $b\simeq a$ used in deriving (\ref{eq:simplemax}), to get:

$$  A\mu \ge 2 a\sqrt{A} + a^2\frac{B}{A}$$

Considering than the goal is to have the smallest possible value of $\mu$ (the minumum detectable signal), and ignoring trivial constants, 
the quantity to be maximized in this case turns out to be:

\begin{equation}\label{eq:multiFOM}
\textrm{F.O.M.} = \frac{A^2}{A^{3/2}+ a B/2 } = \frac{\left[\sum_i{\frac{\epsilon_i^2}{b_i}}\right]^2}{\left[\sum_i{\frac{\epsilon_i^2}{b_i}}\right]^\frac{3}{2}+ \frac{a}{2}\left[\sum_i{\frac{\epsilon_i^3}{b_i^2}}\right]}
\end{equation}

this is a convenient and reasonably simple expression to use in an optimization procedure. Again it has no dependence on the expected signal size, but only on the efficiencies and backgrounds in the individual bins.  It is also easily seen that the expression is invariant for an overall efficiency, multiplying all values of $\epsilon_i$ -- as expected. 

Another interesting form in which the expression can be recast is:

\begin{equation}
\label{eq:multi}
\textrm{F.O.M.} = \frac{A^2/B}{a/2 + A^{3/2}/B}
\end{equation}

because of its similarity in structure with the single-channel formula (\ref{eq:simplemax}). It is easily seen that, for a number of bins $n=1$, this reduces to
 \begin{eqnarray}
A^2/B = \epsilon_1\\
A^{3/2}/B = \sqrt{b_1}
\end{eqnarray}
recovering formula (\ref{eq:simplemax}), that can now be seen simply a special case of (\ref{eq:multi}).

\section{Gaussian case}

A notable and common example of multichannel optimization is given by the case of a gaussian signal superimposed over a slowly varying background.

If we consider a case in which there is a sufficiently large number of events to allow binning the data with a granularity significantly finer that one 
$\sigma$ of the Gaussian, we can address the problem in the frame of the previous section, with individual channel expectations being $\epsilon_i \epsilon(t) N$, where $\epsilon(t)$ is an overall selection-dependent efficiency for the signal, and $\epsilon_i$ simply proportional to the gaussian density function sampled at the bin center:

$$\epsilon_i \propto \exp{\left[-\frac{(x_i-m)^2}{2\sigma^2}\right]}$$

Let's also consider a case in which the the background is sufficiencly slow-varying to be approximable with a flat distribution. For convenience, let us indicate with $\beta(t)$ the number of (selection-dependent) background events in a bin of width $\sigma$. 

With these definitions, and replacing the discrete sum on bins with the integral, we have:

\begin{eqnarray}
A = \int_{-\infty}^{\infty}{\frac{\left(\epsilon(t)\exp{\left[-\frac{(x-m)^2}{2}\right]/\sqrt{2\pi}}\right)^2}{\beta(t)}dx} \\
B = \int_{-\infty}^{\infty}{\frac{\left(\epsilon(t)\exp{\left[-\frac{(x-m)^2}{2}\right]/\sqrt{2\pi}}\right)^3}{\beta(t)^2}} dx
\end{eqnarray}

the two integrals are trivial to calculate, yielding for the two subexpressions appearing in eq. (\ref{eq:multi}):

\begin{eqnarray}
\frac{A^2}{B} = \sqrt{\frac{3}{2}}\epsilon(t)  \\
\frac{A^{3/2}}{B} = \sqrt{\frac{3\sqrt{\pi}}{2}}\sqrt{\beta(t)}
\end{eqnarray}

and finally, dropping irrelevant multiplicative constants:

\begin{equation}
\label{eq:gauss}
\textrm{F.O.M} = \frac{\epsilon(t)}{\frac{a}{2} + \sqrt{\frac{3\sqrt{\pi}}{2}\beta}} \simeq \frac{\epsilon(t)}{a/2 + \sqrt{2.66 \beta(t)}} 
\end{equation}

This is the expression that needs to be maximized when looking for a gaussian signal superimposed over a nearly flat background. 
It is equivalent to expression (\ref{eq:simplemax}) for the simple counting experiment when $B(t)$ is equated to the expected number of background events falling within a $\simeq 2.66 \sigma$ window.

\end{document}